\documentclass[aps,prl,reprint,groupedaddress,superscriptaddress]{revtex4-2}
\usepackage{indentfirst}
\usepackage{amsmath}
\usepackage{graphicx}
\usepackage{dutchcal}
\begin{document}
	
	\title{Synergy between Spin and Orbital Angular Momenta on a Möbius Strip}
	
	\author{Lei Liu}
	\affiliation{National Laboratory of Solid State Microstructures and Department of Materials Science and Engineering, Nanjing University, Nanjing 210093, China}
	\author{Xiao-Chen Sun}
	\affiliation{National Laboratory of Solid State Microstructures and Department of Materials Science and Engineering, Nanjing University, Nanjing 210093, China}
	\author{Yuan Tian}
	\affiliation{National Laboratory of Solid State Microstructures and Department of Materials Science and Engineering, Nanjing University, Nanjing 210093, China}
	\author{Xiujuan Zhang\thanks{Corresponding author}}
	\email[]{xiujuanzhang@nju.edu.cn}
	\affiliation{National Laboratory of Solid State Microstructures and Department of Materials Science and Engineering, Nanjing University, Nanjing 210093, China}
	\author{Ming-Hui Lu\thanks{Corresponding author}}
	\email[]{luminghui@nju.edu.cn}
	\affiliation{National Laboratory of Solid State Microstructures and Department of Materials Science and Engineering, Nanjing University, Nanjing 210093, China}
	\affiliation{Jiangsu Key Laboratory of Artificial Functional Materials, Nanjing 210093, China}
	\affiliation{Collaborative Innovation Center of Advanced Microstructures, Nanjing University, Nanjing 210093, China}
	\author{Yan-Feng Chen\thanks{Corresponding author}}
	\email[]{yfchen@nju.edu.cn}
	\affiliation{National Laboratory of Solid State Microstructures and Department of Materials Science and Engineering, Nanjing University, Nanjing 210093, China}
	\affiliation{Collaborative Innovation Center of Advanced Microstructures, Nanjing University, Nanjing 210093, China}
	\date{\today}
	
	\begin{abstract}
		% insert abstract here
		Spin and orbital angular momenta are fundamental physical characteristics described by polarization and spatial degrees of freedom, respectively. Polarization is a feature of vector fields while spatial phase gradient determines the orbital angular momentum ubiquitous to any scalar field. Common wisdom treats these two degrees of freedom as distinct and independent principles to manipulate wave propagations. Here, we demonstrate their synergy. This is achieved by introducing two orthogonal $p$-orbitals as eigenbases, whose spatial modal features are exploited to generate orbital angular momenta and the associated orbital orientations provide means to simultaneously manipulate polarizations. Through periodic modulation and directional coupling, we realize a full cyclic evolution of the synchronized and synergized spin-orbital angular momenta. Remarkably, this evolution acquires a nontrivial geometric phase, leading to its representation on a Möbius strip. Experimentally, an acoustic cavity array is designed, whose dipole resonances precisely mimic the $p$-orbitals. The acoustic waves, uniquely, see the pressure (scalar) field as a spatial feature and carry an intrinsic polarization defined by the velocity (vector) field, serving as an ideal platform to observe the synergy of spin and orbital angular momenta. Based on such a property, we further showcase a spin-orbital-Hall effect, highlighting the intricate locking of handedness, directionality, spin density and spatial mode profile. Our study unveils a fundamental connection between spin and orbital angular momenta, promising avenues for novel applications in information coding and high-capacity communications.
	\end{abstract}

	%\keywords{}
	
	\maketitle
	\textit{Introduction}.---Exploring the intricate interplay between spin (SAM) and orbital angular momentum (OAM) has been a longstanding and compelling quest in wave physics. Spin is an intrinsic form of angular momentum carried by elementary particles like electrons\cite{merzbacher1998quantum}. Over a century ago, Poynting predicted a circularly polarized light carries angular momentum\cite{poynting1909wave}, nowadays attributed to the $\hbar$ spin of the photons, known as SAM. In a broader context, SAM characterizes the orientations of the electric or magnetic fields, i.e., the polarizations (the vector aspects of the electromagnetic waves)\cite{varshalovich1988quantum}. Apart from SAM, light also carries OAM, which was identified by Allen et al. in 1992\cite{allen1992orbital}. They recognized that unlike SAM being more like an intrinsic characteristic of light, OAM can be generated by twisted azimuthal phase gradients which are associated with spatial degrees of freedom and ubiquitous in all types of waves. Benefited from such a property, OAM has been vastly explored in various realms of waves, including structured light\cite{forbes2021structured}, quantum optics\cite{lodahl2017chiral}, light manipulations based on meta-structures (metamaterials and metasurfaces)\cite{yu2011light,song2021plasmonic,ahmed2022optical}, and even longitudinal waves like acoustic waves\cite{jiang2016convert,fu2022asymmetric,ge2023spatiotemporal}.
	
	While SAM and OAM are often treated as independent principles governing distinct degrees of freedom (SAM for the vector aspects and OAM for the spatial scalar aspects), they can interact with each other, pertaining to the so-called spin-orbital interactions (SOIs)\cite{bliokh2015spin}. There are primarily two types of SOIs. The first one is associated with the helicity-dependent position or momentum for wave propagations, known as the spin-Hall effect\cite{bliokh2015quantum}. One famous example is the classical version of the topological quantum spin-Hall effect. Therein, the propagations of the topological edge states are found to depend on the helicity of SAM\cite{kane2005quantum,bernevig2006quantum}. The second type of SOI concerns the combination of the vector features of SAM with different orders of OAM, expanding scalar OAM waves to more general types of vectorial vortices\cite{milione2011higher,naidoo2016controlled}. In these vectorial states, the vector polarization constantly changes direction when tracing the azimuthal phase gradient, suggesting a SAM-OAM conversion. Despite different features in SOIs, the essence is the discussion on how one angular-momentum component affects the other.
	
	Historically, SAM and OAM can be independently manipulated or interact with each other. This forms the traditional understanding. However, a crucial piece has been missing—whether SAM and OAM can be synchronized or synergized. Here, we bridge this gap by demonstrating the synergy of SAM and OAM on a Möbius strip, which exhibits a full cyclic evolution from linear-polarization into circular-polarization and back to itself, accompanied by changes in OAM from linear momentum to a perfect vortex and back. The revelation of such a fundamental and unique connection introduces additional degrees of freedom to manipulate waves, with high potential in information coding and high-capacity communications.
	\begin{figure}[htbp]
		\centering
		\includegraphics[width=1\linewidth]{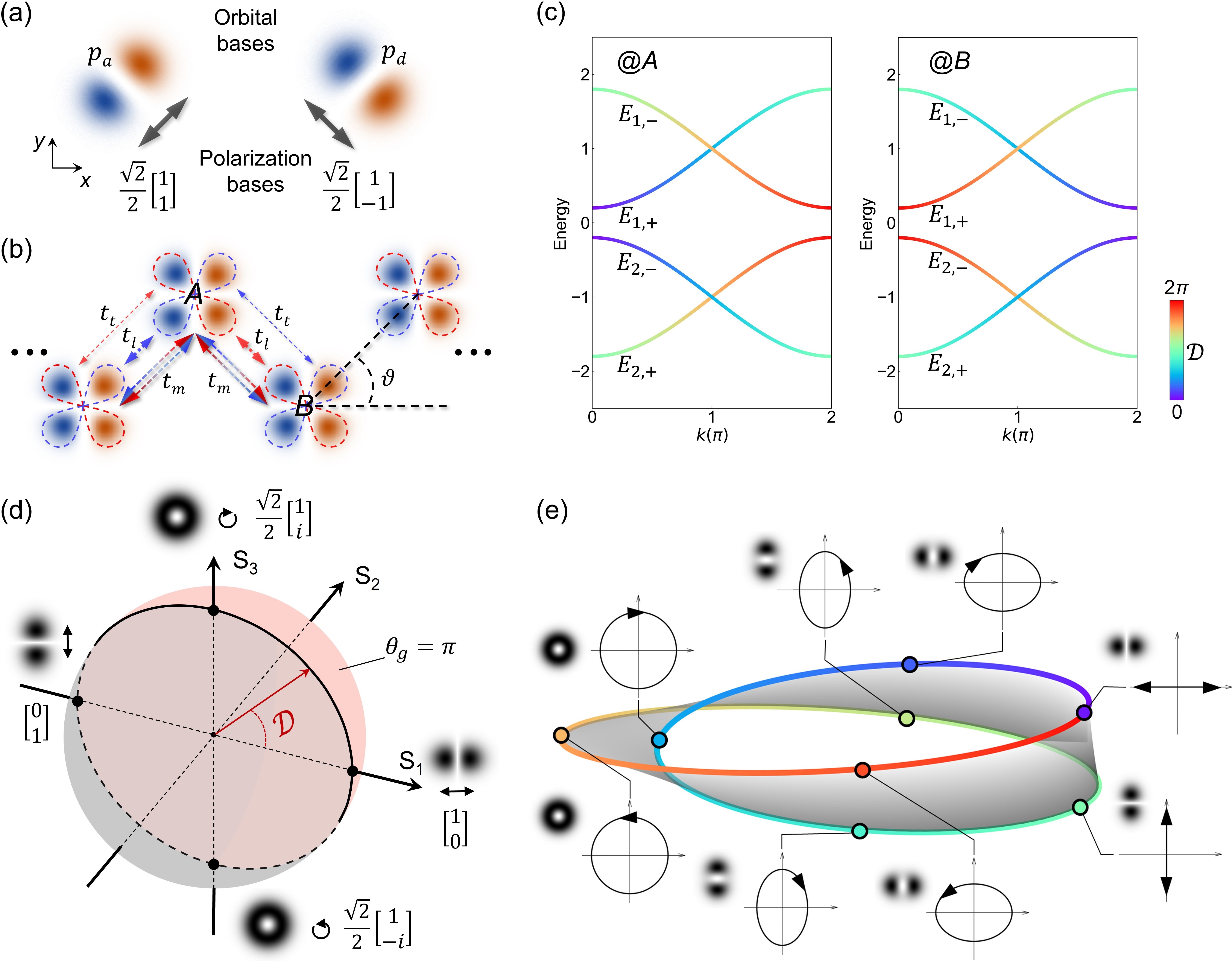}
		\caption{\label{fig1}
			(a) Schematics of two orthogonal $p$-orbitals.			
			(b) A 1D lattice constructed by periodically arranging the $p$-orbitals.		
			(c) Band structures of the lattice in (b), incorporating $\mathcal{D}$. The left panel is for Site A and the right for Site B.			
			(d) One cyclic evolution of the eigenstates in (c) by tracing $\mathcal{D}$ from $0$ to $2\pi$, mapped onto a Poincaré sphere and exemplified for Site A in the $E_1$ group. As the cyclic evolution acquires a Pancharatnam-Berry geometric phase $\theta_g=\pi$, it can be further represented on a Möbius strip as shown in (e) where both the OAM states and polarization ellipses are presented.}
	\end{figure} 

	\textit{Synergy between SAM and OAM}.---We start with two orthogonal $p$-orbitals, labelled by $|p_a\rangle$ and $|p_d\rangle$, as schematically illustrated in Fig. \ref{fig1}(a). Without loss of generality, the orientation of $|p_a\rangle$ ($|p_d\rangle$) is set along the antidiagonal (diagonal) direction. In the Jones vector representation, the corresponding vector bases are ${\frac{\sqrt2}{2}\left[\begin{matrix}1&1\\\end{matrix}\right]}^T$ and ${\frac{\sqrt2}{2}\left[\begin{matrix}1&-1\\\end{matrix}\right]}^T$. The orbitals are arranged into a quasi-one-dimensional (1D) lattice along the $x$-direction, as depicted in Fig. \ref{fig1}(b). Each unit cell comprises two sites, labeled by A and B, each hosting a pair of orthogonal $|p_a\rangle$ and $|p_d\rangle$. Sites A and B are further dislocated along the $y$-direction to introduce a directional coupling (characterized by the angle $\vartheta$), which effectively generates distinct gauge potentials for $|p_a\rangle$ and $|p_d\rangle$, leading to a phase difference between them and therefore giving rise to nonzero OAM. Concurrent with the same phase change, the vector bases are superposed, leading to nonzero SAM in synergy with the OAM. Due to the periodicity, the phase change depends on the wave vector and is periodic as well. Correspondingly, the OAM and SAM undergo a full cycle of synergy.
	
	The proposed model is characterized by three coupling coefficients, i.e., $t_t$, $t_l$ and $t_m$ for the transverse, longitudinal and crossed coupling, respectively (see illustrations in Fig. \ref{fig1}(b)). For this model, the Hamiltonian is written as $H(k)=\begin{bmatrix} \mathbf{0}_{2\times2} & h(k) \\ h^\dag(k) & \mathbf{0}_{2\times2} \end{bmatrix}$, with $\ h(k) = (t_t + t_l e^{ik})\sigma_0 + (t_m + t_m e^{ik})\sigma_x$ ($\sigma_0$ and $\sigma_x$ are the Pauli matrices), corresponding to eigenfunction $|\psi\rangle=(\phi_{A,p_a},\phi_{A,p_d},\phi_{B,p_a},\phi_{B,p_d})^T$. Here, $\phi_{i,p_j}$ with $i=A,B$ and $j=a,d$ represents the $|p_j\rangle$ component on Site $i$, the lattice constant is assumed to be $1$, $k$ denotes the wave vector, and $\dag$ indicates the complex conjugate transpose. 
	
	The phase difference between $|\left.p_a\right\rangle$ and $|\left.p_d\right\rangle$ components is obtained as $\mathcal{D} = \arg{\left(\frac{\phi_{i,p_a}}{\phi_{i,p_d}}\right)}$. To track the evolution of $\mathcal{D}$, we present the energy bands in Fig. \ref{fig1}(c) taking $\vartheta=60^{\circ}$, $t_t=0$, $t_l=-1$ and $t_m=0.4$, where $\mathcal{D}$ is visualized through color-coding (more details can be found in Supplementary Material\cite{supp}, which includes Refs. \cite{ge2021observation,caceres2020topological,schulz2022photonic,barnett2001optical,vanderbilt2018berry,slobozhanyuk2015subwavelength,dennis2009singular,gong2018nanoscale}). There are four bands grouped in two, denoted by $E_1$ and $E_2$ with $+$ and $-$ signs representing positive and negative group velocity, respectively. In the $E_1$ group, $\mathcal{D}$ undergoes an evolution from $0$ to $2\pi$, with one half following the $E_{1,+}$ band and the other following the $E_{1,-}$ band. In the $E_2$ group, the evolution is symmetric to the $E_1$ group with respect to the zero-energy. For different Sites A and B, $\mathcal{D}$ exhibits opposite evolutions. These observations elucidate the intricate dependence of $\mathcal{D}$ on $k$, which is intimately tied to the symmetries of the system\cite{supp}. 
	
	Associated with the evolution of $\mathcal{D}$, the OAM and SAM exhibit a full cycle of synergy. To visualize the cyclic evolution, we employ the Poincaré sphere representation\cite{born2013principles,padgett1999poincare}. Typically, a polarized state (or an OAM state) can be described on a sphere spanned by the Stokes parameters $\{S_1, S_2, S_3\}=\{\cos{2\varphi} \cos{2\chi}, \sin{2\varphi} \cos{2\chi}, \sin{2\chi}\}$, with $2\varphi$ and $2\chi$ respectively the azimuthal and ellipticity angles. In our scenario, the OAM and polarization states are superpositions of the $|\left.p_a\right\rangle$ and $|\left.p_d\right\rangle$ orbital bases and their orientation vector bases, yielding $|\left.p_d\right\rangle + e^{i\mathcal{D}}|\left.p_a\right\rangle$ and $\frac{\sqrt{2}}{2}\left(\begin{bmatrix} 1 & -1 \end{bmatrix}^T + e^{i\mathcal{D}}\begin{bmatrix} 1 & 1 \end{bmatrix}^T\right)$, respectively. Such a choice of eigenbases pins the evolution onto the $S_1$-$S_3$ plane (corresponding to $\varphi=0$).
	
	A full cycle of evolution can be tracked by $\mathcal{D}=2\chi$ running over $[0, 2\pi]$, as illustrated in Fig. \ref{fig1}(d), exemplified for Site A in the $E_1$ group. Initially, at $E_{1,+}(k=0)$, the OAM state is expressed as $|\left.p_d\right\rangle + |\left.p_a\right\rangle$, corresponding to a $p$-orbital with orientation along the $x$-direction. This state can be mapped to the east of the equator of the Poincaré sphere, in synergy with a horizontal linear-polarization. As $k$ increases, the linear-polarization transforms into elliptical-polarizations, associated with distorted $p$-orbitals. Upon reaching the poles, the evolution state becomes a perfect vortex and is simultaneously circularly-polarized. For different sites, the evolution occurs in opposite directions, resulting in vortices with topological charges of $l=\pm1$ and circular-polarizations with opposite handedness, i.e., $|\left.p_d\right\rangle + i|\left.p_a\right\rangle$ and $\frac{\sqrt{2}}{2}e^{i\frac{\pi}{4}}\begin{bmatrix} 1 & i \end{bmatrix}^T$ for Site A at the north pole and $|\left.p_d\right\rangle - i|\left.p_a\right\rangle$ and $\frac{\sqrt{2}}{2}e^{-i\frac{\pi}{4}}\begin{bmatrix} 1 & -i \end{bmatrix}^T$ for Site B at the south pole\cite{supp}. Continuous increase in $k$ leads to the evolution from the north pole to the west of the equator for Site A and from the south pole to the west of the equator for Site B. Correspondingly, the OAM state returns to the $p$-orbital and the polarization to the linear-polarization, only with a vertical orientation.This completes half of the evolution. By following the $E_{1,-}$ band, the other half unfolds, precisely reversing the first half of A-evolution for Site B and the first half of B-evolution for Site A. In this way, a full cycle of SAM-OAM synergy is realized. For the $E_2$ group, a similar analysis can be performed.
	
	It is known that a cyclic evolution on a Poincaré sphere generates a nontrivial Pancharatnam-Berry geometric phase $\theta_g = \Omega/2$, where $\Omega$ is the solid angle enclosed by the cyclic evolution\cite{pancharatnam1956generalized,berry1984quantal}. In our case, the cyclic evolution encloses half of the Poincaré sphere, giving rise to $\theta_g = \pi$. Such nonzero geometrical phase in the momentum space indicates nontrivial topology of the periodic lattice\cite{supp}. It is also suggested that the eigenstates can be represented on a Möbius strip with $4\pi$ periodicity\cite{manoharan2010romance,xue2022projectively,li2022acoustic}. This is consistent with our observation, i.e., the eigenstates return to their initial states only after $4\pi$ change of $k$ (or $2\pi$ change of $\mathcal{D}$). Figure \ref{fig1}(e) illustrates how one cycle of the SAM-OAM synergy is represented on the Möbius strip.
	
	Simultaneously manipulating anisotropic orbitals and their orientations opens up opportunities to synchronize OAM and SAM, uncovering a fundamental connection between these angular momenta. We argue that such a principle is universal and can be extended to a general selection of orbitals and orientations, offering versatile control over the synergy between OAM and SAM\cite{supp}.
	\begin{figure}[htbp]
		\centering
		\includegraphics[width=0.9\linewidth]{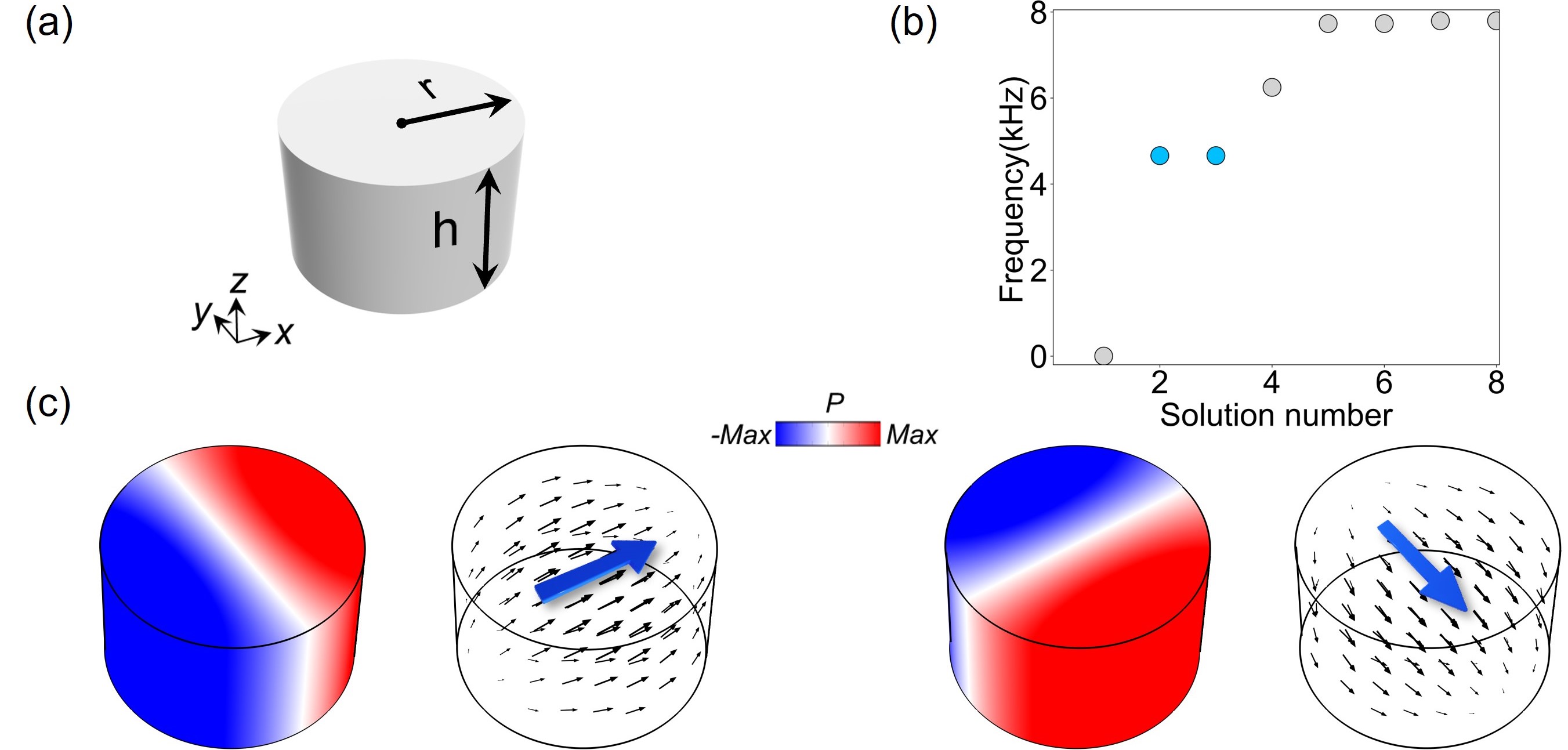}
		\caption{\label{fig2}
			(a) Schematics of an acoustic cavity. 
			(b) Resonant spectra of the cavity with two degenerate dipole resonances marked by the blue dots.
			(c) Distributions of acoustic pressure (color map) and velocity (thin arrows) fields of the two dipole modes. The orientations of the modes are indicated by two thick arrows.
		}
	\end{figure}
	\begin{figure*}[htbp]
		\centering
		\includegraphics[width=0.9\linewidth]{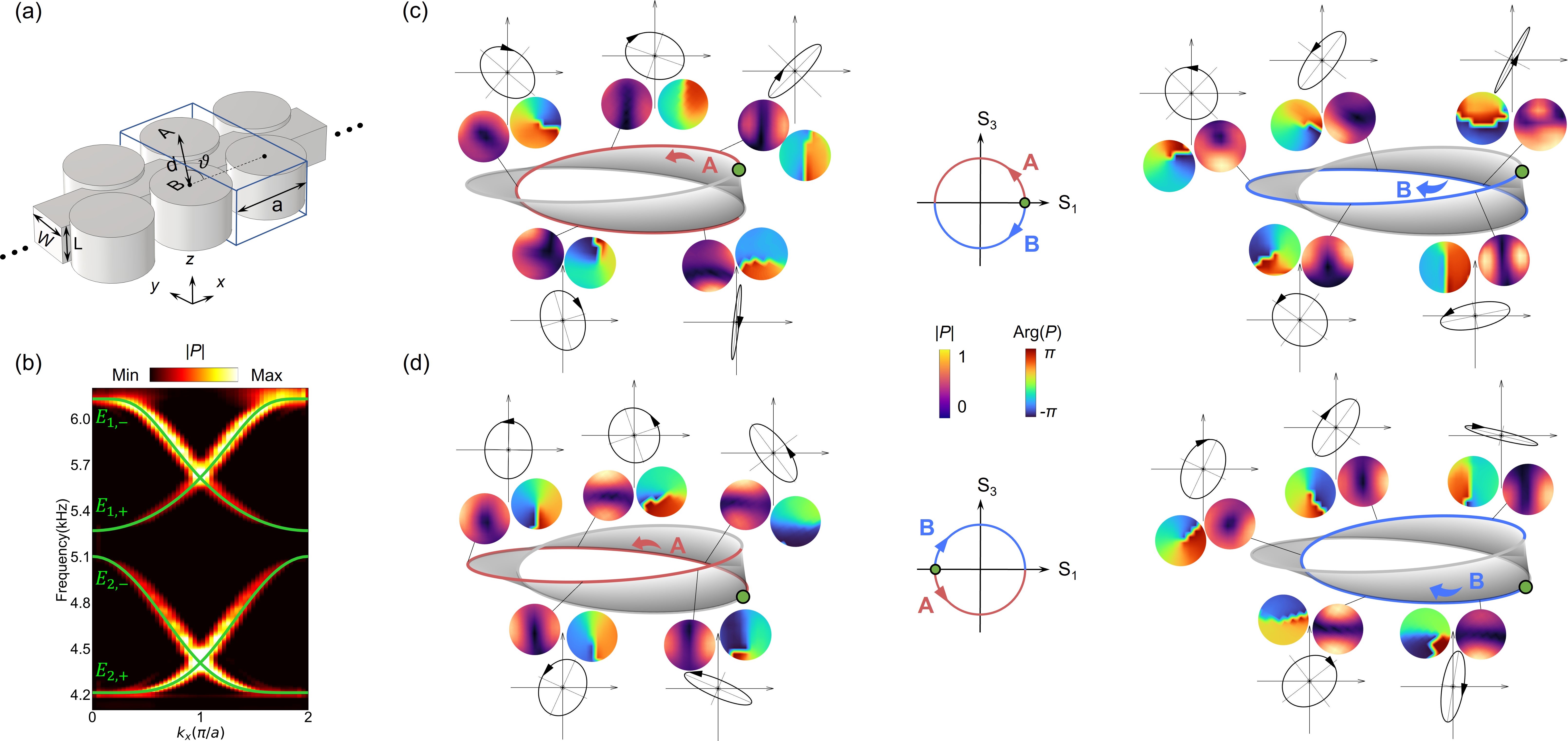}
		\caption{\label{fig3}
			(a) Schematics of the quasi-1D acoustic lattice. The blue lines frame a unit cell. 
			(b) Numerically calculated (green curves) and experimentally measured (color map) band structures for the lattice in (a). 
			(c) Measured acoustic pressure fields $|P|$, their phase distributions $\mathrm{Arg}(P)$, and polarization ellipses of velocity fields for cavities A and B following $E_{1,+}$ evolution. Both representations on the Möbius strip and indication on the $S_1$-$S_3$ plane of the Poincaré sphere are provided. 
			(d) The same as (c), only for $E_{2,+}$ evolution.}
	\end{figure*}
	
	\textit{Experimental demonstration}.---Applicable as a general principle, the revelation of synergy between SAM and OAM is particularly noteworthy for scalar waves such as acoustic waves, which are conventionally considered polarization-free. Here, we show acoustic waves, with their pressure fields as spatial degrees of freedom and velocity fields as vector degrees of freedom, inherently embody the conditions outlined in our theoretical framework. This positions them as an ideal platform for demonstrating the SAM-OAM synergy and exploring the associated physical phenomena.

	We consider an acoustic cylindrical cavity illustrated in Fig. \ref{fig2}(a), which supports two degenerate dipole resonances, as shown by the eigenspectra and field distributions in Figs. \ref{fig2}(b-c) (see Ref.\cite{supp} for geometric and material parameters). Their pressure fields precisely mimic the spatial features of two orthogonal $p$-orbitals and the velocity fields exactly align with the orientations of the $p$-orbitals. Such a correspondence is inherently rooted in the incompressible linear Euler equation where the vector velocity field is connected with the gradient of the scalar pressure field\cite{pierce2019acoustics}.
	
	To observe the SAM-OAM synergy, we further construct a 1D periodic array of the acoustic cavities connected via a rectangular air channel, as illustrated in Fig. \ref{fig3}(a). The band structures are presented in Fig. \ref{fig3}(b), both numerically calculated and experimentally measured (see details in Ref.\cite{supp}), which exhibit a remarkable agreement with the energy bands in Fig. \ref{fig1}(c). The cyclic evolution and SAM-OAM synergy along the $E_{1,+}$ band are displayed in Fig. \ref{fig3}(c), where the pressure fields and their phase distributions are measured as spatial OAM features and the polarization ellipses are presented based on the measurements of the velocity fields, as vector polarization features. Tracking the A-evolution, it is observed to precisely agree with the theoretical prediction, i.e., from the east of the equator passing the north pole to the west of the equator, an $x$-oriented $p$-orbital (accompanying the horizontal linear-polarization) transforms into a vortex with $l=+1$ (accompanying the right-circular-polarization), then returns to a $y$-oriented $p$-orbital (accompanying the vertical linear-polarization). For the B-evolution, it is also from an $x$-oriented $p$-orbital to a $y$-oriented $p$-orbital, only passing a vortex with $l=-1$ (at the south pole).

	Similar to Fig. \ref{fig3}(c), Fig. \ref{fig3}(d) presents the cyclic evolution along the $E_{2,+}$ band. We point out that due to the technical difficulties at the band edges (with zero group velocity), our measurements only cover the $k$ regime close to $[0,\ 2\pi]$. Nonetheless, the measurements evidently reflect the fundamental features of the cyclic evolutions with high agreements with the theory (see Ref.\cite{supp} for simulations covering a complete $k$ regime).
	
\begin{figure}[htbp]
		\centering
		\includegraphics[width=1\linewidth]{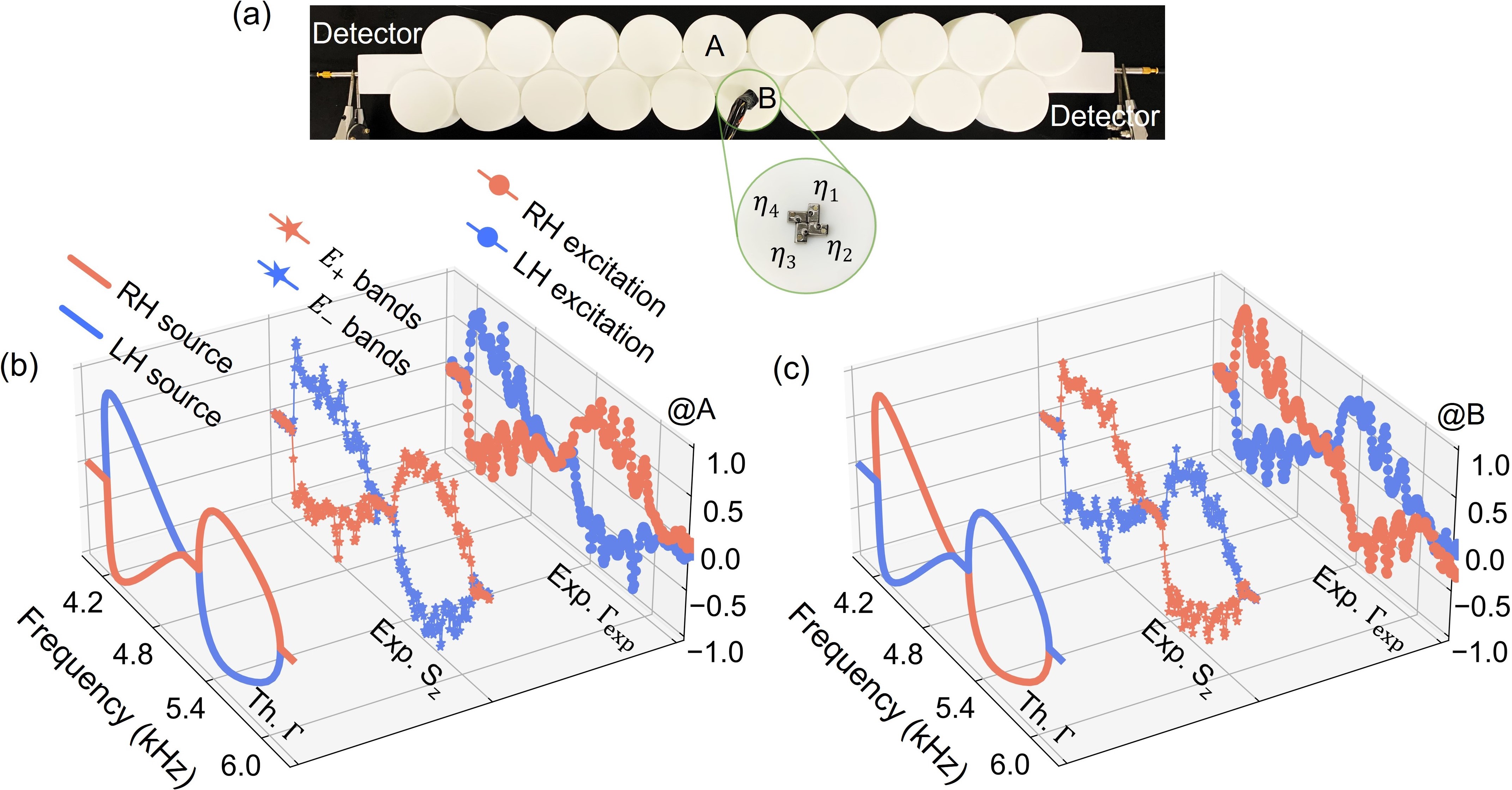}
		\caption{\label{fig4}
			(a) The experiment setup.
			(b) Numerically calculated directional contrast $\Gamma$ (left panel), experimentally measured spin density $S_z$ (middle panel), and directional contrast $\Gamma_{\text{exp}}$ (right panel) as functions of frequencies, for Site A.
			(c) The same as (b), only for Site B.
		}
	\end{figure}

	\textit{A spin-orbital-Hall effect}.---As a fundamental property enabled by SOIs, SAM can be locked with the linear momentum\cite{bliokh2015spin}. In our case, remarkably, SAM is synchronized with the OAM, promising a novel spin-orbital-Hall effect highlighting more interesting and intricate locking of handedness, directionality, spin density and spatial mode profile. For demonstration, an acoustic sample is fabricated as shown in Fig. \ref{fig4}(a). To probe the locking properties, we employ a chiral source generated by coupling four acoustic loudspeakers with a phase gradient\cite{supp}. This kind of source carries nonzero angular momentum that can couple with the SAM and correspondingly with the synchronized OAM, resulting in an SAM-OAM-dependent sound propagation.
	
	Such dependence can be quantified by a directional contrast $\Gamma = \frac{\Gamma_f - \Gamma_b}{\Gamma_f + \Gamma_b}$, with $\Gamma_f$ and $\Gamma_b$ describing the likelihood of the wave propagation in the forward and backward directions, respectively. According to Fermi’s golden rule\cite{grynberg2010introduction,coles2016chirality,lang2022perfect}, $\Gamma_f \propto |\mathbf{s}^* \cdot \mathbf{v}_f|^2$ and $\Gamma_b \propto |\mathbf{s}^* \cdot \mathbf{v}_b|^2$\cite{supp}. Here, $\mathbf{s} = \frac{\sqrt{2}}{2} \begin{bmatrix} 1 & \pm i \end{bmatrix}^T$ denote the chiral sources carrying SAM with opposite handedness. $\mathbf{v}_f$ and $\mathbf{v}_b$ indicate the velocity fields for eigenstates with positive and negative group velocity, respectively. Physically, this determines how much of the energy in the source $\mathbf{s}$ is transferred into the polarization states defined by $\mathbf{v}_f$ and $\mathbf{v}_b$.
	
	Figures \ref{fig4}(b)-(c) present $\Gamma$ calculated based on the band structures in Fig. \ref{fig3}(b), for Sites A and B, respectively. It is seen that a right-handed chiral source with $\mathbf{s}=\frac{\sqrt{2}}{2}\begin{bmatrix}1&i\end{bmatrix}^T$ (denoted by red lines) transfers energy into Site A (B) as a forward (backward) propagating state in the high-frequency regime, but as a backward (forward) propagating state in the low-frequency regime. When the source chirality is reversed, with $\mathbf{s}=\frac{\sqrt{2}}{2}\begin{bmatrix}1&-i\end{bmatrix}^T$ (denoted by blue lines), the sign of $\Gamma$ flips. This suggests that the right- (left-) handedness couples to the polarization states with a preference in the north (south) hemisphere, which is reasonable given the polarization features analyzed in the cyclic evolutions.

	Enabled by the cyclic evolutions, more interestingly, the polarization-dependent wave propagation is tunable with respect to different frequencies. Tracking the frequency axis in both Figs. \ref{fig4}(b)-(c), we see that the $\Gamma$-curves display two maxima around 4.4 kHz and 5.6 kHz, precisely corresponding to the poles accommodating circular-polarizations. At these points, the source energy is maximally transferred into a single mode, leading to unidirectional sound propagation. Away from the maxima, the source energy is split, with a large portion transferred into the preferable direction and a small portion into the opposite direction with reversed handedness. The transition ratio is quantified by $\Gamma$. In fact, such tunable polarization-splitting is related to the change of spin density during the cyclic evolutions. The spin density is defined as the angular momentum carried by the velocity fields, yielding $\mathbf{S}=\frac{\rho}{2\omega}\mathrm{Im}(\mathbf{v}^\ast\times\mathbf{v})$\cite{long2018intrinsic,bliokh2019spin}. It characterizes the acoustic polarizations of local velocity field rotations, with normalized $\mathbf{S}$ being zero representing linear-polarizations, unity representing circular-polarizations (positive for the right-circular-polarization and negative for the left-circular-polarization) and in between representing elliptical-polarizations. Using the experimental data measured in Fig. \ref{fig3}, we calculate $S_z$ ($S_x=S_y=0$ due to zero $v_z$). As presented in Figs. \ref{fig4}(c)-(d), $S_z$-curves are consistent with the $\Gamma$-curves (with oscillations from the finite size effect).
	
	Remember that in our system, the SAM and OAM are synchronized and synergized, suggesting a tunable OAM-splitting, accompanying the tunable polarization-splitting. For experimental demonstration, we place two detectors at the right and left ends of the sample in Fig. \ref{fig4}(a) to measure the output pressure signals in the forward ($P_f$) and backward ($P_b$) directions, respectively. An experimental directional contrast is obtained as $\Gamma_{exp}=\frac{|P_f|-|P_b|}{\left|P_f\right|+|P_b|}$ and plotted in Figs. \ref{fig4}(b)-(c). Again, $\Gamma_{exp}$-curves exhibit high agreements with the calculated $\Gamma$ and measured $S_z$, indicating the spatial OAM modes indeed exhibit a tunability in synergy with the polarization states (see more details in Ref.\cite{supp}).
	
	\textit{Conclusions}.---We have demonstrated a full cycle of synergy between SAM and OAM on a Möbius strip by exploiting anisotropic $p$-orbitals as eigenbases, whose spatial mode profiles and inherent orientations serve as ingredients to simultaneously manipulate OAM and SAM. The uncovering of such a unique and fundamental connection deepens our understanding of these angular momenta, which traditionally are considered independent and separate. It is especially prominent for scalar waves like acoustic waves, which are revealed to naturally carry synchronized SAM and OAM due to the intrinsic connection between scalar pressure and vector velocity fields. The current model with simple acoustic cavities already glimpses into the rich dynamical physics of the SAM-OAM synergy and tunable SAM-OAM-locked wave propagations. Facilitated with artificial designs, more comprehensive models taking into account various physical principles like topology, gauging, and pumping would promise versatile wave controls. While investigating $p$-orbitals in one dimension, our principle can be applied to a more general selection of orbitals, orientations, and spatial dimensions, with more degrees of freedom for rich SAM-OAM interplay. In addition, synchronized SAM and OAM feature non-separable states imprinting simultaneously spatial scalar and vector signatures with high potential in information coding and high-capacity communications. Targeting on the on-chip information technologies, our principle may offer an integrated solution for coupled SAM-OAM control with reduced fabrication complexity.
	\begin{acknowledgments}
		This work is supported by the Key R\&D Program of Jiangsu Province (Grant No. BK20232015), the National Natural Science Foundation of China (Grant No. 12222407), and the National Key R\&D Program of China (Grants No. 2023YFA1407700 and No. 2023YFA1406904). X.Z. thanks Z. Zhang for helpful discussions.
	\end{acknowledgments}
	\bibliography{references}
\end{document}